# Two-dimensional structure reconstruction with expectation and maximization algorithm

Yun Zhao

Department of Physics, Arizona State University, Tempe, Arizona 85287, USA

**Abstract**

In this report, we applied expectation and maximization (EM) method described by Philips et al [1] to recover two-dimensional (2D) structure from multiple sparse signal images in random orientation. The detailed derivation of EM algorithm for 2D image reconstruction was evaluated. Data sets with average 40 photons per frame were successfully classified by orientation. And the 2D mask structure is reconstructed by merging all frames with the appropriate rotations applied to each one. It provides us an alternative approach in data set classification and structural information recovery from extremely weak signal with incomplete information.

## 1 Introduction

Much efforts have been devoted to study the structure of single particle with x-ray free electron laser(XFEL), which could produce very intense femtosecond x-ray pulse[2][3]. This new method could potentially overcome the radiation damage on crystals as well as other limitations on traditional techniques[2][3][4]. However, each diffraction snapshot collected from single particle contains very few photons, as the interaction between single particle and x-ray is too weak[4]. An intuitive solution is through merging all the snapshots to obtain the complete diffraction pattern. The problem stems from that we can't tell the orientation of particle just by each snapshot nor by direct observation. And particles will be destroyed during each shot. The difficulty is even exacerbated as the existence of background radiation noise. So a fundamental question in front of us is whether we are able to distinguish the orientation of two noisy diffraction patterns with sparse photons in principle.

One approach to classify the orientation is based on cross-correlation method by Huldt et al.[5]. They successfully classified the diffraction patterns with around one photon per pixel. However, the photon fluence in our scenario is about 0.001 photons per pixel, much lower than Huldt's case. And the cross-correlation method would fail in the ultra-low fluence limit[1]. Another robust method addressed to solving sparse randomly-oriented x-ray data was based on expectation-maximization(EM) method. EM method was first introduced to find parameters for a statistical model with incomplete data in information theory. Elser is one of the earliest people introduced this method in structure reconstruction from sparse randomly-oriented data[6][7].

In this report, we focused on a 2D object with 4 random orientations during imaging,

a much simpler case where I believe it is more illustrative to show the principle and feasibility of EM algorithm in structure reconstruction. Nobody could reconstruct structure from single snapshot with only one photon. Here we explore the minimum requirement for photons fluence to recover structure with given number of frames. Noise effects were also discussed. A detailed evaluation of EM algorithm for image reconstruction is also given the following parts.

## 2   An intuitive explanation of EM algorithm

In general, EM method seeks to finding some unknown parameters of statistic model by iteration, given measurement data which contains some unobserved variables[8]. Here is an outline of EM iteration.

Given a statistical model consisting of a set of observed data X, with missing values Z. We may start a random guess for unknown parameters $\theta$. Then the likelihood function could be expressed as $L(\theta; X, Z) = p(X, Z|\theta)$. And the maximum likelihood estimate of the unknown parameters is determined by the marginal likelihood of the observed data

$$L(\theta; X) = p(X|\theta) = \sum_Z p(X, Z|\theta)$$

The iteration could be described as following two steps[8]:
E-step: Estimate the expectation value of log-likelihood function, given distribution Z with parameter $\theta^{(i)}$ in ith iteration.

$$Q(\theta^{(i+1)}|\theta^{(i)}) = \langle \log(L(\theta^{(i)}; X)) \rangle = \langle \log(\sum_Z p(X, Z|\theta^{(i)})) \rangle$$

M-step: Determine the new $\theta^{(i+1)}$ which could maximize $Q(\theta^{(i+1)}|\theta^{(i)})$.

The parameter $\theta$ will converge to a optimal value by iteratively applying the above two steps.

One of the earliest paper on EM algorithm is given by Hartley in 1958 [9]. In that paper, he simplified the procedure for seeking the maxi-mum likelihood computations of estimates from incomplete data by iteration. The iteration idea was also generated to several cases. However, the EM algorithm was first explicitly explained and given its name by a classic paper by Dampster, Laird and Rubin[8]. They formulated EM algorithm by defining expectation and maximization step during each iteration and generated its application to a wider class of statistical models. In particular, they also gave rigorous proof for the convergence of EM iterations for several models. More details on the convergence of EM algorithm can also be found in a book by G. McLachlan, and T. Krishnan [10].

# 3 Methods and Results

## A. Data collection

The experiment designed here is almost the same as the one described in Philipp[1]. We simulated the imaging process of a 2D L-shape mask with extreme weak signals. The rotation of mask is spaced by 90°. The detector in our simulation is a 200×200 pixel array. The orientation of mask will be reset randomly in one of the four equally possible orientation after an image is taken. Data sets with different quality are obtained by changing the photon counts per frame recorded during simulation. 10 000 snapshots were generated for each case.

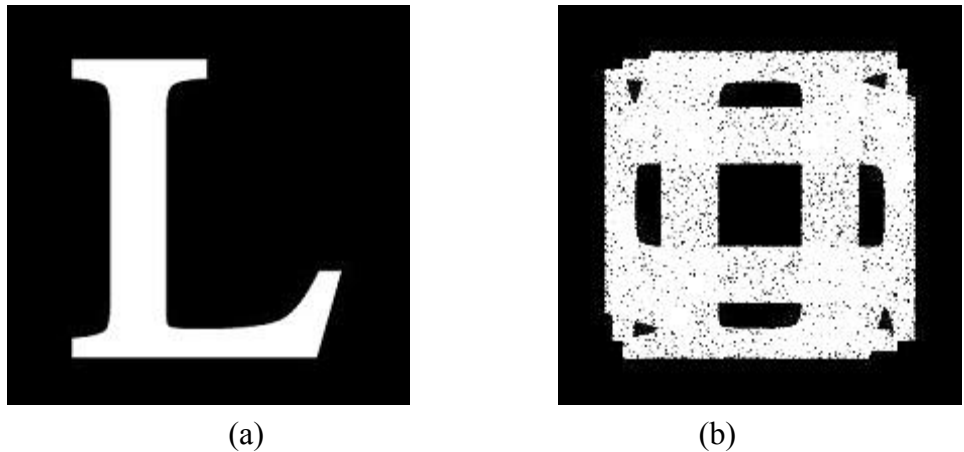

(a)　　　　　　　　　　　　(b)

Fig 1. (a) The L-shape mask with a square aperture. (b) Sum of all frames with 40 photons per frame data set, showing a uniform distribution with 4 possible orientations.

## B Image reconstruction with EM algorithm

The algorithm we have adopted for the image reconstruction is based on the idea of expectation maximization. My interpretation here is largely based on several papers[6][8][9][11] and a book[10]. The derivation and idea are almost the same as Elser's work on reconstruction algorithm[1][6]. Here I just gave more details and interpreted in a slightly different perspective, which perhaps easier to understand.

The parameter in the present setting is the intensity signal model $w$, a 200×200 matrix. The data collected are the sets of frames with photon counts $k$ recorded by the detector, where the orientation of the mask relative to the detector $r$ is intractable. Our model is updated, $w \rightarrow w'$, based on maximizing a log-likelihood function $Q(w')$. While orientation probability distribution of each frame $p_{rf}$ is based on the current model parameters $w$. As we have 10 000 frames and 4 possible orientations, so $p_{rf}$ is a 10 000×4 matrix in our algorithm.

Let's use $w_r$ to denote the intensity distribution on detector when the image is in rotation r. The $f$th snapshot is assigned a probability distribution, $p_{rf}$, with respect to its unknown rotation, r, relative to the current intensity model. The rotations are sampled in increments of $2\pi/N$, where $N$ defines the angular resolution of the reconstruction. $N$ is 4 in our case as we know the number of possible orientations in imaging process in advance. Each frame comprises photon occupancy, $k_{if}$, at pixel $i$, which in our low-fluence experiment are almost zero, the exceptions being equal to 1. Because the photon counts are independent Poisson samples of the intensity at each pixel, the probability is

$$p_{rf} \propto \prod_i \frac{w_{ir}}{k_{if}!} e^{-w_{ir}} \propto \prod_{i \in I_f} w_{ir}$$

where $ir$ is rotation $r$ applied to pixel $i$, $I_f$ is the set of pixels recording photons in frame $f$.

Then the probability of f_th frame in orientation $r$ could be normalized by

$$p_{rf} = \frac{\prod_{i \in I_f} w_{ir}}{\sum_r \prod_{i \in I_f} w_{ir}}$$

Note here that the probability is calculated by the current model $w$.

The log-likelihood function for f_th frame in orientation $r$ is

$$Q_{rf}(w') = \log \left( \prod_i \frac{w_{ir}'^{k_{if}}}{k_{if}!} e^{-w_{ir}'} \right)$$

$$= \sum_{i=1}^{N_{pix}} \log \left( \frac{w_{ir}'^{k_{if}}}{k_{if}!} e^{-w_{ir}'} \right)$$

$$= \sum_{i=1}^{N_{pix}} (k_{if} \log w_{ir}' - w_{ir}' - \log k_{if}!)$$

As $k_{if} = 1$ or $0$, so $k_{if}! = 1$ and $\log k_{if}! = 0$.

$$Q_{rf}(w') = \sum_{i=1}^{N_{pix}} (k_{if} \log w_{ir}' - w_{ir}')$$

Now the expectation of log-likelihood function may now be written explicitly:

$$Q(w'|w) = \sum_{f=1}^{N_{frame}} \sum_{r=1}^{N_{rot}} P_{rf}(w) Q_{rf}(w')$$

$$= \sum_{f=1}^{N_{frame}} \sum_{r=1}^{N_{rot}} \sum_{i=1}^{N_{pix}} (P_{rf}(w) k_{if} \log w_{ir}' - P_{rf}(w) w_{ir}')$$

After obtaining the expectation estimate for $Q(w'|w)$, the algorithm proceeds to the second step.

$w'$ is obtained by solving the equation $\frac{dQ(w')}{dw'} = 0$, as it should maximize the value of $Q(w')$. Note that $P_{rk}(w)$ comes from expectation step, which depends on the

current model w, rather than new model w′. So the maximizing update rule is given by

$$w_{ir} \to w'_{ir} = \frac{\sum_{f=1}^{N_{data}} P_{rf}(w)k_{if}}{\sum_{f=1}^{N_{data}} P_{rf}(w)}$$

Note that $w_{ir}$ is the intensity of pixel i when the mask in orientation r. At the last step, we merge the models from different orientations.

$$w'(i) = \langle \sum_r p_{rf} k_{-rf} \rangle_f$$

where −rf means a rotation applied on frame f in the opposite direction of r.

The updated intensity model w′ is an average of the photon counts in all frames with the appropriate distribution of rotations applied to each one. Each element in w′ will be very tiny number after averaging, as each frame contains very few photons. In practical simulation, we need to amplify our final model w′ by multiplying a proper constant to obtain a bright image, or it will be very dark.

**4 Results and Discussion**

The EM iteration starts from a random model with each element assigned a random number in the range of [0,1], as shown in Fig 2a. At the end of iteration, the model will end up a structure with arbitrary orientation. Figure 2a was reconstructed using 10 000 frames of data with an average of 40 photons per frame. This data set has a total of 0.5 million photons. For comparison, a data set with the same total frame but a higher photon fluence was also processed. The reconstruction is shown in Fig 2d, where the average occupancy was 150 photons/frame.

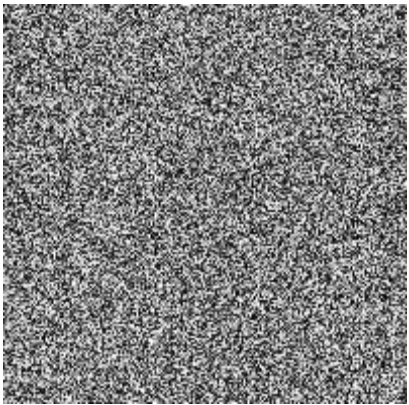 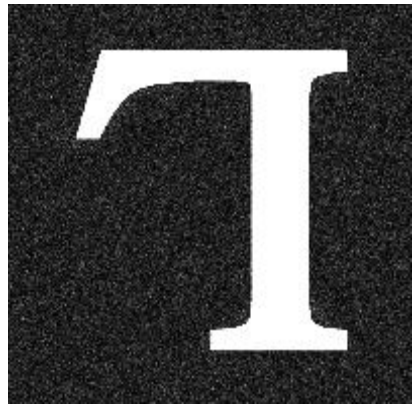

(a)                                (b)

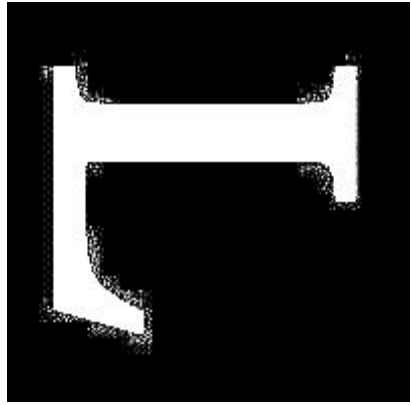 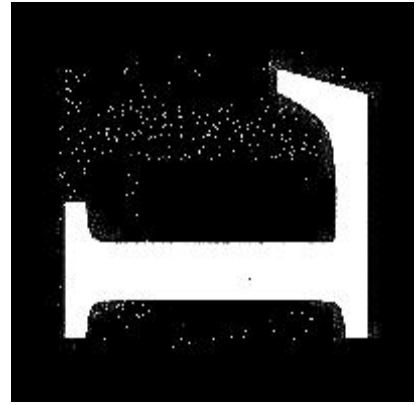

(c)　　　　　　　　　　　　　(d)

Fig 2 (a) initial random model with no structural information. (b) A reconstruction using random-oriented data having a average 150 photons/frame with S/N=10; (c) A reconstruction using random-oriented data having a average 40 photons/frame; (d) A reconstruction using random-oriented data having a average 150 photons/frame.

The quality of the two reconstruction differs in classification accuracy, with the 150 photons/frames data yielding better results. There is a also increase in the iteration count of the EM algorithm: the 40 photons/frame data required 32 iterations, compared with only 4 iterations for the 150 photon/frame data. The minimum requirement for photon fluence is 40 photons/ frame, which is much higher than 2.5 photon/frame in Philipps' paper. This difference mainly comes from the fact that they have a much larger data set with 450 000 frames, which is 45 times bigger than mine.

Images with noise are also studied here. Here we assume the background radiation is incoherent and uncorrelated between pixels. And the net signal is simply the sum of x-ray scattering from mask as well as background. And the S/N is defined as the average signal matrix element over the average noise matrix element. A successful reconstruction for a average 150 photons/frame with S/N=10 data set was shown in fig 2b. The presence of noise degrades the image quality and raises the requirement for minimum photon fluence. The longest simulation I made is for 50 photons per frame data set with $S/N = 1$. It took for 287 iterations without any sign showing convergence.

In addressing noise problem, Elser gave the criteria for different classification methods[7]. In that paper, he proposed that the arbitrarily high level of noise could be tolerated as long as an unlimited measurements are available[7].

The EM algorithm demonstrated above could be generated to 3D reconstruction[6][11]. In that scenario, the 3D intensity model will be expanded into tomographic representation at first as the information recorded by our detector is 2D information. This work was already done by Loh et al. And their for a 3D particle reconstruction is available on line[12].

The last comment I'd like to make is about the limitation about the algorithm. The theoretical model matrix is pretty much binary as all the elements could just be 1or 0, white or black in our image. And our approximation in $p_{rf}$ estimation is greatly based on this assumption. If the elements in a model could be any real number between 0 and 1, can we still recover the model? The above algorithm failed to reconstruct it even with thousands photons per frames. An possible solution is that we just give up the approximation for Poisson distribution $\prod_i \frac{w_{ir}}{k_{if}!} e^{-w_{ir}} \propto \prod_{i \in I_f} w_{ir}$. But it will be much more computational expensive. As to this aspect, cross-correlation method seems to play a complementary role in addressing this problems.

## 5 Conclusion

The motif of this study was to demonstrate the principle of EM algorithm in sparse signal image reconstruction and classification. The minimum requirement for successful structure recovery depends on the photon fluence per frame, size of data set as well as S/N ratio. By comparison the simulation presented here with Philip's work[1], it seems that the minimum requirement for photon fluence can be relieved by producing a larger data set.

**Acknowledgement**
The author thanks Prof. John Spence for discussion.

**References and links**
1   H. T. Philipp, K. Ayyer, M. W. Tate, V. Elser, and S. M. Gruner, "Solving structure with sparse randomly-oriented x-ray data," OPTICS EXPRESS 20, 13129 (2012).
2   J C H Spence, U Weierstall and H N Chapman, "X-ray lasers for structural and dynamic biology," Rep. Prog. Phys. 75 (2012) 000000 (25pp)
3   R. Neutze, R. Wouts, D. van der Spoel, E. Weckert, and J. Hajdu, "Potential for biomolecular imaging with femtosecond x-ray pulses," Nature406, 752–757 (2000).
4   R. Fung, V. Shneerson, D. K. Saldin and A. Ourmazd, "Structure from fleeting illumination of faint spinning objects in flight," Nature Physics 5, 64–67 (2008)
5   G. Huldt,  A. Szoke, and J. Hajdua, " Diffraction imaging of single particles and biomolecules" Journal of Structural Biology 144 219–227 (2003)
6   N.-T. D. Loh and V. Elser, "Reconstruction algorithm for single-particle diffraction imaging experiments," Phys. Rev. E80, 026705 (2009).
7   V. Elser, "Noise limits on reconstructing diffraction signals from random tomographs," IEEE Trans. Inf. Theory55, 4715–4722 (2009).
8   A.P. Dempster, N.M. Laird, D.B. Rubin, "Maximum Likelihood from Incomplete Data via the EM Algorithm". Journal of the Royal Statistical Society. Series B (Methodological) 39 (1): 1–38. (1977).
9   H. Hartley, "Maximum likelihood estimation from incomplete data," Biometrics,14:174–194. (1958).

10    G. McLachlan, and T. Krishnan, The EM algorithm and extensions. Wiley series in probability and statistics. John Wiley & Sons. (1997).
11    N. D. Loh, M. J. Bogan, V. Elser, A. Barty, S. Boutet, S. Bajt, J. Hajdu, T. Ekeberg, F. R. N. C. Maia, J. Schulz, M. M. Seibert, B. Iwan, N. Timneanu, S. Marchesini, I. Schlichting, R. L. Shoeman, L. Lomb, M. Frank, M. Liang, and H. N. Chapman, "Cryptotomography: reconstructing 3D Fourier intensities from randomly oriented single-shot diffraction patterns," Phys. Rev. Lett.104, 225501 (2010).
12    http://www.duaneloh.com/